# Measuring Space-Time Geometry over the Ages

**Albert Stebbins**
Theoretical Astrophysics Group
MS209
Fermi National Accelerator Laboratory
Box 500
Batavia, IL 60510
*stebbins@fnal.gov*

# Abstract

Theorists are often told to express things in the "observational plane". One can do this for space-time geometry, considering "visual" observations of matter in our universe by a single observer over time, with no assumptions about isometries, initial conditions, nor any particular relation between matter and geometry, such as Einstein's equations. Using observables as coordinates naturally leads to a parametrization of space-time geometry in terms of other observables, which in turn prescribes an observational program to measure the geometry. Under the assumption of vorticity-free matter flow we describe this observational program, which includes measurements of gravitational lensing, proper motion, and redshift drift. Only 15% of the curvature information can be extracted without long time baseline observations, and this increases to 35% with observations that will take decades. The rest would likely require centuries of observations. The formalism developed is exact, non-perturbative, and more general than the usual cosmological analysis.

A major goal of observational cosmology over the past century has been the determination of the large scale geometry of space-time. The values of $H_0$ and $\Omega_0$ are still important topics of research. [1,2] More recently cosmologists have pursued the characterization of a repulsive force causing the expansion of the universe to accelerate. [3] This force is believed to be gravitational, which in modern parlance means that it is a result of curvature of space-time. We study gravity/geometry by observing objects like galaxies which follow geodesics (apart form other *known* forces) and using photons which also follow geodesics. It is this geodesic motion which measures the geometry.

Newton first determined the inverse square nature of gravitational forces by measuring acceleration in the trajectories of planets. Modern cosmologists have a very different toolkit. Our tools include the Hubble diagram (the relationship between luminosity distance, $D_L$, and redshift, $z$; gravitational lensing shear (the distortion of shapes of objects by the gravity along the line-of-sight to these objects), and various features of the large scale structure (LSS) ($\equiv$ inhomogeneities). [3] What is surprising about this list is that none of these, with the exception of shear, measures acceleration of anything: LSS measures positions, the Hubble diagram measures distance versus velocity (redshift). This is kinematics *not* dynamics!

The reason these methods can be used to learn about gravitational forces is because of the various assumptions that are used in the standard cosmological paradigm, most importantly homogeneity and isotropy of the universe and initial conditions. Can we determine the geometry of space-time ($\equiv$gravity) more directly? The goal is to only use what you see (observe) to infer what you get (gravity-geometry), rather than the standard approach which involves both seeing and believing ($\equiv$assuming). This is not to say that nothing is assumed, but rather many fewer assumptions are made. In particular we do not assume the Cosmological Principle, general relativity, nor even isotropy about the observer. The formalism works for general space-times and is exact and non-perturbative.

Ideally one would measure geometry by using yardsticks and clocks however in cosmology we are restricted to observations from a single vantage point and over a fairly short period of time. The standard observational paradigm, which focuses on measuring the observed properties of distant objects is used here. The objects are referred to as "matter" which is only required to be observable and freely falling.

To describe a space-time one usually chooses space-time coordinates. Various types of coordinates are popular with theorists, but observers also use coordinates and they have very little freedom in what they can choose. They record the time of observation of an object, $t$; redshift, $z$; and two angular coordinates, denoted $\vartheta^a$, giving position on the sky. It is assumed the observer is comoving with the matter, freely falling and non-rotating. Since we live in an expanding universe cosmologists commonly use this "redshift space" representation of space-time since it is a nearly 1-to-1 representation of the observable part of universe (there are workarounds for when it is not). These coordinates define observationally meaningful submanifolds: past light cones (null 3-surfaces of constant $t$), redshift spheres (2-surfaces of constant $t$ and $z$), and lines-of-sight (null curves of constant $t$, $z$, and $\vartheta^a$).

In redshift space, $\{t, z, \vartheta^a\}$, the matter 4-velocity is

$$u^\mu = \frac{1}{1+z} \begin{pmatrix} 1 \\ \dot{z} \\ \dot{\vartheta}^a \end{pmatrix} \tag{1}$$

where $\dot{z}$ and $\dot{\vartheta}^a$ gives the redshift drift [4,5] and proper motion [6] of matter, *i.e.* the (observer) time rate of change of the redshift, and angular position. It is assumed that $u^\mu$ is vorticity-free, as is common in cosmology. Since the matter is freely falling the vorticity will remain zero if it is zero at some initial time. Also

$$u_\mu = -c^2 \, T_{;\mu} \tag{2}$$

where along any matter trajectory $T$ gives the progression of proper time. $T$ allows one to compare time throughout space-time, and can be tied to observer time by $T_{z=0} = t$. These equations plus requiring lines-of-sight to be null geodesics yield a metric of the form

$$ds^2 = -c^2 \, dT^2 + c^2 \left( dT - \frac{dt}{1+z} \right)^2 + (d\boldsymbol{s}_\perp - \boldsymbol{v}_\perp \, dt) \cdot (d\boldsymbol{s}_\perp - \boldsymbol{v}_\perp \, dt) \tag{3}$$

where $d\boldsymbol{s}_\perp$ and $\boldsymbol{v}_\perp$ are 2-vectors on the redshift spheres and

$$dT \equiv \dot{T} \, dt + T' \, dz + T_{,a} \, d\vartheta^a \qquad \boldsymbol{v}_\perp \cdot d\boldsymbol{s}_\perp \equiv \dot{\vartheta}^a \, D_{ab} \, d\vartheta^b$$

$$d\boldsymbol{s}_\perp \cdot d\boldsymbol{s}_\perp \equiv d\vartheta^a \, D_{ab} \, d\vartheta^b \qquad \boldsymbol{v}_\perp \cdot \boldsymbol{v}_\perp \equiv \dot{\vartheta}^a \, D_{ab} \, \dot{\vartheta}^b \qquad . \tag{4}$$

$$\dot{f} \equiv \frac{\partial f}{\partial t} \qquad f' \equiv \frac{\partial f}{\partial Z} \qquad f_{,a} \equiv \frac{\partial f}{\partial \vartheta^a}$$

One can think of $d\boldsymbol{s}_\perp$ as the incremental transverse distance and $\boldsymbol{v}_\perp$ as the matter transverse velocity. $D_{ab}$ is the metric of the redshift spheres and $D_{ab}$'s inverse can be written

$$D^{ab} = \frac{\hat{\gamma}_+^a \hat{\gamma}_+^b}{D_+{}^2} + \frac{\hat{\gamma}_-^a \hat{\gamma}_-^b}{D_-{}^2} \tag{5}$$

$$\hat{\gamma}_\pm^a \, \gamma_{ab} \, \hat{\gamma}_\pm^b = 1 \qquad \hat{\gamma}_\pm^a \, \gamma_{ab} \, \hat{\gamma}_\mp^b = 0$$

where $D_- \geq D_+ > 0$, $\hat{\gamma}_\pm^a$ form a position angle basis, and $\gamma_{ab}$ is the angular metric such that $d\vartheta = \sqrt{d\vartheta^a \, \gamma_{ab} \, d\vartheta^b}$ is the angular increment in radians. For curves on the sky with tangent $\hat{\gamma}_\pm^a$ one finds that $\frac{d\text{length}}{d\vartheta} = D_\pm$ on the redshift sphere. The angular diameter distance and reduced shear [7] are given by

$$D_{\mathrm{A}} = \sqrt{D_+ D_-} > 0 \quad \text{and} \quad \Gamma = \frac{1}{2} \left( \frac{1}{D_+{}^2} - \frac{1}{D_-{}^2} \right) \geq 0 \tag{6}$$

and $\hat{\gamma}_+^a$ is the position angle of the shear [a].

The observer coordinates dictates the observables that should be measured in order to quantify the geometry: lensing, $D_{\mathrm{A}}$, $\Gamma$ and $\hat{\gamma}_+^a$; proper motions, $\dot{\vartheta}^a$; and time, $T$. $D_{\mathrm{A}}$ is related to the (bolometric) luminosity distance in the Hubble diagram by $D_A = D_L (1 + z)^2$. What is missing is good methodologies for measuring $T$. $T$ is related to the line-of-sight matter velocity gradient, $H_\parallel$ by $T' = -H_\parallel^{-1}$, so one could measure $H_\parallel$ instead. $T$ is also related to redshift drift by

$$\dot{T} + \dot{z}\, T' + \dot{\vartheta}^{'a}\, T_{,a} = \frac{1}{1+z}. \tag{7}$$

but measuring $\dot{z}$ is not sufficient to determine $T$. Methodologies to obtain $T$ or $H_\parallel$ include cosmochronology, absolute volume measures, and Alcock-Pacynski tests [8]. Let us proceed assuming an accurate methodology will arise. If it does not then we do not know the line-of-sight distance scale which largely precludes any localized measure of curvature.

There is continuing improvement in astrometry. The upcoming Gaia [9] mission and a proposed OBSS [10] mission expect to achieve, $\pm 200\mu\text{asec}$ and $\pm 10\mu\text{asec}$, astrometry for $10^9$ stars. Galaxy proper motions, are in this range $\sim \frac{10\,\mu\text{asec}}{\text{decade}} \frac{v_\perp}{500\,\text{km/sec}} \frac{100\,\text{Mpc}}{(1+z)\,D_A}$. Redshift drift is also small, but in a few decades adds up to the velocity precision obtained for stars in planet searches ($10^{-9}\,c$). Admittedly both $\dot{z}$ and $\dot{\vartheta}^{'a}$ are more difficult for galaxies than for stars, but the signal grows with time. Improving technology will eventually push uncertainties below the increasing signal. We will likely be doing this science sometime this century.

A geometer is most interested in the Riemann curvature which is a localized measure of geometry. An Einstein believing physicist is most interested in the Ricci curvature which is a localized measure of the stress-energy. These curvatures depend on gradients of the metric. Redshift and angular gradients can be done by differencing between adjacent object in redshift space. Temporal gradients require patience. Observations which can be done "right away" include lensing and hopefully $T$ and are accessible to the *impatient cosmologist*. Measurements of $\dot{\vartheta}^{'a}$ and $\dot{z}$ may take decades and so we say they are accessible to a *patient cosmologist*. Time derivatives of lensing or of $\dot{\vartheta}^{'a}$ or of $\dot{z}$ will take even longer and are relegated *very patient cosmologists* and their descendants. Second time derivatives will take even longer and are relegated to *foolish cosmologists*. The longer time period are difficult to estimate since technology will improve and better methodologies may develop.

Construct an orthonormal tetrad, $\{\hat{u}^\mu, \hat{r}^\mu, \hat{v}^\mu, \hat{w}^\mu\}$, where in the matter frame $\hat{u}^\mu$ is the time direction , $\hat{r}^\mu$ is the line-of-sight direction, $\hat{v}^\mu$ is the proper motion direction, and $\hat{w}^\mu$ is what's left. Define the null vector $\hat{n}^\mu \equiv \hat{r}^\mu - \hat{u}^\mu$. In the $\{\hat{u}^\mu, \hat{n}^\mu, \hat{v}^\mu, \hat{w}^\mu\}$ basis the patience level (I,P,V,F for impatient, patient, very patient, foolish) of Ricci tensor components are

$$\begin{pmatrix} R_{uu} & R_{un} & R_{uv} & R_{uw} \\ R_{nu} & R_{nn} & R_{nv} & R_{nw} \\ R_{vu} & R_{vn} & R_{vv} & R_{vw} \\ R_{wu} & R_{wn} & R_{wv} & R_{ww} \end{pmatrix} \sim \begin{pmatrix} F & V & V & V \\ V & I & P & P \\ V & P & V & V \\ V & P & V & V \end{pmatrix} \tag{8}$$

and of Riemann tensor components are

$$
\begin{array}{llll}
R_{unun} \sim P & R_{unuv} \sim V & R_{unuw} \sim V & R_{unnv} \sim P \\
R_{unnw} \sim P & R_{unvw} \sim V & R_{uvuv} \sim F & R_{uvuw} \sim F \\
R_{uvnv} \sim V & R_{uvvw} \sim V & R_{uwuw} \sim F & R_{uwnv} \sim V \\
R_{uwnw} \sim V & R_{uwvw} \sim V & R_{nvnv} \sim I & R_{nvnw} \sim I \\
R_{nvvw} \sim P & R_{nwnw} \sim I & R_{nwvw} \sim P & R_{vwvw} \sim V
\end{array}
\tag{9}
$$

(the rest can be determined by symmetries). Only 3 of 10 components of the Ricci curvature and 7 of 20 components of the Riemann tensor are accessible to the patient cosmologist, and only 1 of 10 and 3 of 20 to the impatient cosmologist. One would have to develop different methodologies or introduce additional assumptions to reduce the wait.

By now the reader is probably impatient to hear more about $R_{nn}$. The expression for $R_{nn}$ is a form of the optical Raychaudhuri equation [11], but in terms of observables. In the context of Einstein's equations it gives $T_{nn}$ which is totally insensitive to a cosmological constant. Patient readers would be interested to know that for an isotropic fluid comoving with the matter $R_{nv} = R_{nw} = 0$ and gives no additional information about the equation-of-state.

To summarize, it was shown how to build space-time geometry purely out of observations by a single freely-falling observer in an expanding universe. The resulting formalism is exact and involves no assumptions about isometries or the relationship between matter and curvature, *e.g.* Einstein's equations. The lack of assumptions limits the inferential power of the observations, but makes any results much more general. This formalism will be presented in more detail elsewhere [12].

**Acknowledgement:** I would like to thank Scott Dodelson, Lam Hui, Rocky Kolb, Umeh Obina, and Robert Wald for useful discussions. This work was supported by the DOE at Fermilab under Contract No. DE-AC02-07CH11359

## Footnotes

*a.* The relationship between $D_{ab}$ and $D_A$ and $\Gamma$ would be more obvious if we were considering surfaces of constant $T$ rather than constant $z$. However one can always boost to a frame where the surface of constant time is tangent to the redshift sphere. Since Lorentz boosting an object neither magnifies nor distorts it's image, but rather makes it appear rotated, it does not change the angular diameter distance $D_A$ or the reduced shear $\Gamma$. Any surface in a past light cone will have the same relationship between it's metric and the lensing parameters.